# Third-generation cylindrical dendrimers based on L-aspargic acid in solutions: hydrodynamic and electrooptical properties

Ilya Martchenko and Nikolai Tsvetkov

Department of Physics, St Petersburg State University
Ul. Ulyanovskaya, 3, 198504 Stary Peterhof, Russia

*ilyamartch@mail.ru* *n.tsvetkov@paloma.spbu.ru*

Samples of third-generation cylindrical dendrimers with molar masses ranging in the interval *20000... 60000* have been studied by the methods of equilibrium and non-equilibrium electrical birefringence, molecular hydrodynamics and optics. It was found that the absolute values of Kerr and flow birefringence constants exceed the values obtained for analogous dendrimers of lower generations. The mechanism of reorientation has proven to be strongly dependent on the physical and chemical properties of the solvent. In chloroform solutions, the studied dendrimers align to the microwave-frequency electric fields according to large-scale mechanism. In dichloroacetic acid solutions, the observed reorientation mechanism is low-scale, which is explained by degradation of intermolecular hydrogen bonds. Terminal dendritic substituents of the macromolecules have experimentally proven to be oriented mainly along the primary polymer chain.

## Introduction

There have been growing interests in the macromolecules that contain branched side substituents and seriously differ from chain polymers. If the structure of a synthesized repeatedly branched molecule is regular, it is called *dendrimer* (Greek: δενδρον, meaning *tree*.) The possibility of building dendrimers was first predicted by Flory in 1951 [1], and the synthesis strategies have been developed by Wögtle [2], Tomalia [3] and Newkome [4] in 1970s – 1980s. A notable contribution to the study and systematization of dendrimers was made by Muzafarov [5].

Dendrimers are classified into several structural types, including *spherical dendrimers* (with branched groups attached to a single central core) and *cylindrical* or *linear dendrimers* (with branched groups attached to a linear polymer chain) [6]. Number of cascade-like ramifications determines the generation of a dendrimer (G1, G2, G3…). A serious attention has been recently paid to analyzing the difference between dendrimers built of flexible and of rigid internal units [7].

Profoundly researched and typically monodisperse spherical dendrimers have proven to approach continuous non-draining particles in terms of their hydrodynamic properties. Much less publications have focused on cylindrical dendrimers. It is known that the direction of the main polymer chain determines a selected axis in these macromolecules. The equilibrium rigidity of cylindrical dendrimers is typically moderate. The conformation of a macromolecule, its molar mass and polymerization degree, also its physical properties such as persistent length, axial ratio, kinetic elasticity of a chain and possible helix conformation are not always well controlled at synthesis and may be determined *a posteriori* with physical methods.

A great interest in recent years is devoted to dendrimers that include amino acids as their main chains [8] or side dendrons [9, 10]. Such polymers may find applications in pharmaceutics and in genetical research because of their probable biological activity.

In present work, samples of third-generation cylindrical dendrimers with molar masses ranging in the interval *20000...60000* have been studied by methods of electrical birefringence (in rectangular-pulsed and sinusoidal-pulsed electric fields) and molecular hydrodynamics and optics.

Structure of a monomer unit (with molar mass of *1656*) is outlined on Fig. 1.

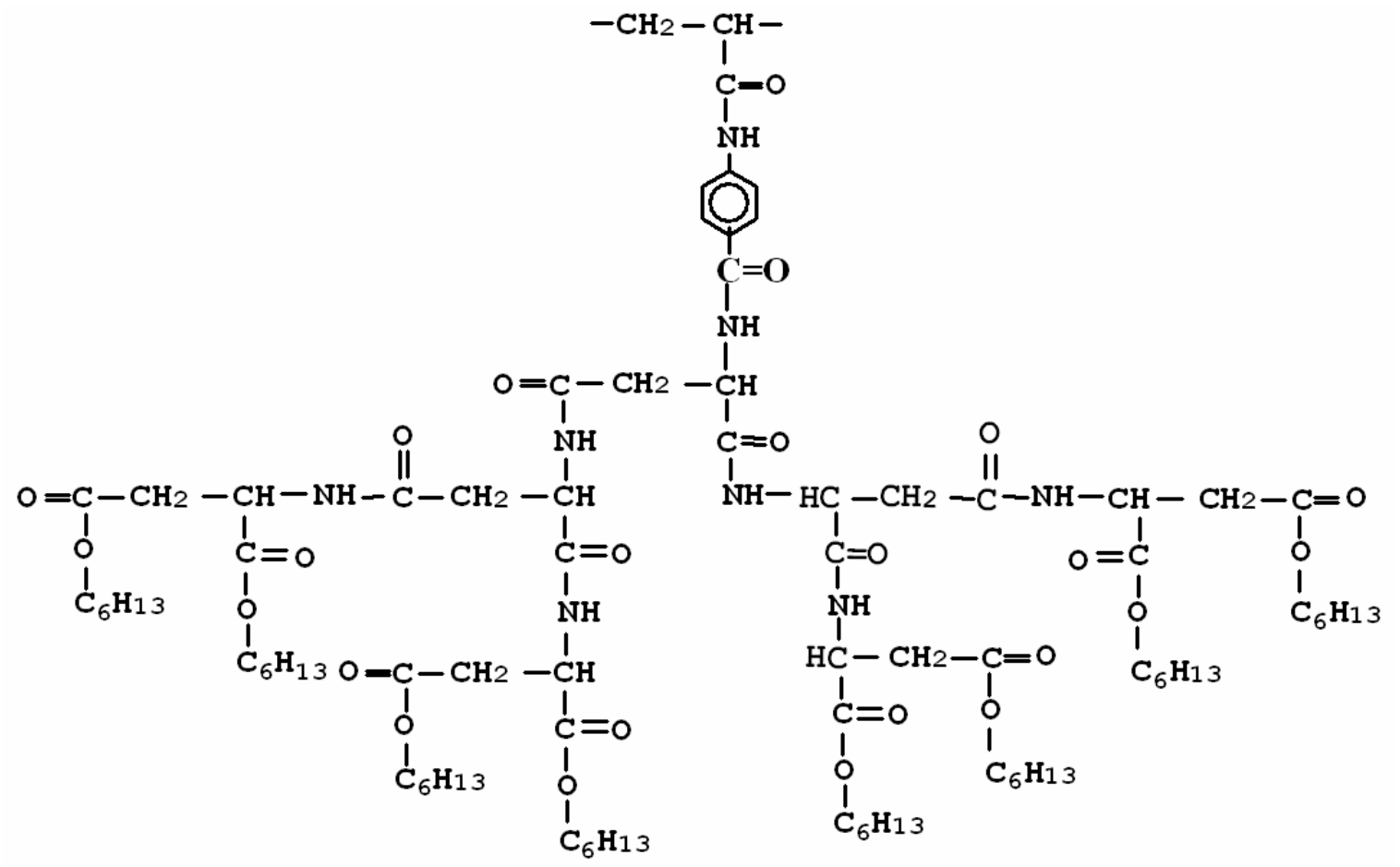

Such a molecule experiences multiple intermolecular hydrogen bonds. It also includes an anisotropic aromatic fragment and long aliphatic groups $C_6H_{13}$ which increase its solubility in organic solvents. The synthesis of samples has been performed by the macromolecular compound division at the Chemistry Department, St Petersburg State University.

All samples varied only in their polymerization degree. Solutions of analogous dendrimers of first and second generations have been studied earlier [11, 12, 13, 14, 15].

# Methods

## Equilibrium electric birefringence

Kerr has experimentally determined that the optical anisotropy induced in an isotropic medium by external electric field is proportional to the squared amplitude of the field. For solutions, Kerr constant is introduced as the characteristic of an infinitely diluted solution in an infinitely weak field:

$$K = \lim_{c \to 0, E \to 0} \frac{\Delta n}{cE^2},$$

where $\Delta n$ is the difference of refractive indexes for ordinary and extraordinary beams, $c$

is the mass concentration of the solved material (*g/cm*$^3$).

Electric birefringence may be of both positive and negative signs, and is determined by the magnitude of molecule's dipole moment and the angle between the vector of dipole moment and the symmetry axes of optical and dielectric polarizabilities of a molecule. Therefore, an analysis of equilibrium birefringence allows discussing not only the dipole moment, but also the conformation of a macromolecule.

Equilibrium electric birefringence was studied in rectangular-pulsed electric fields. Duration of pulses was *3-5 ms* at the frequency of 1 pulse per sec. The use of pulsed fields was necessary for eliminating parasite effects not relevant to changes of optical anisotropy (such as heating or ion drift.)

A Kerr cell equipped with two titanium electrodes was filled with the studied solution. The gap between the electrodes was equal to $d=0,030\pm0,005$ *cm*, the length of the electrodes along the light path was equal to *3 cm*. A He-Ne laser that generated the wavelength $\lambda=632.8$ *nm* was used as a light source. In order to improve sensitivity, an elliptic rotating compensator with the relative path difference of $\Delta\lambda/\lambda=0.01$ was used.

The compensation method has serious advantages in comparison to photoelectric methods that register the radiant intensity. The secondary optical effects, caused by electric field, modulate the radiation intensities while the optical anisotropy remains constant. The compensation method involves selectors and elliptic modulators and allows sensitively detecting the changes in the anisotropy of solution, but not the secondary modulations of light.

An oscilloscope was used to register the photoelectric current generated by photomultiplier. The magnitude and the sign of the electric birefringence were determined with the rotational displacement of the compensator that was used to compensate incident optical pulses detected by photomultiplier. Calibration was performed against the data for benzene which has known Kerr constant.

**Non-equilibrium electric birefringence**

A momentary delay (or lag) needed for molecules to align in sinusoidal-pulsed electric field is determined by their rotational diffusion coefficients.

The frequency dispersion of electric birefringence helps to obtain direct information on the relaxation spectra and rotational mobility of a macromolecule. The specific Kerr constant at a fixed frequency of sinusoidal-pulsed electric field is given by:

$$K_\upsilon = \frac{\Delta n - \Delta n_0}{cE^2},$$

where $\Delta n$ is the birefringence at given frequency, $\Delta n_0$ is the birefringence of a solvent, $c$ is the mass concentration.

The equilibrium and non-equilibrium measurements have been performed with similar equipment.

**Flow birefringence**

The flow birefringence is a dynamical orientation of non-spherical molecules in a laminar flow. This method permits to directly find the optical anisotropy and rotational diffusion coefficients of macromolecules and helps to effectively study their spatial conformation.

A titanium dynamooptimeter with the height of *3.21 cm* and the rotor's diameter of *3 cm* was filled with the solution. The gap between the concentric cylinders was *0.022 cm*. A He-Ne laser that generated the wavelength $\lambda=632.8\ nm$ was used. The elliptic rotating compensator had the path difference $\Delta\lambda/\lambda=0.036$. The measurements were made with a photoelectric sensor. To avoid temperature effects, the apparatus was thermostatted at 24 $^0$C with water bath. Since the measurements were made at small shear stresses, there was made an assumption that the angles between the optical axes of solutions and the flow directions were close to 45$^0$.

The optical shear coefficient $\Delta n/\Delta\tau$ is given by:

$$\frac{\Delta n}{\Delta \tau} = \frac{\Delta n_p - \Delta n_0}{g(\eta - \eta_0)},$$

where $\eta$ is the viscosity of a solution, $\eta_0$ is the viscosity of a solvent, $\Delta n_p$ is the birefringence of a solution, $\Delta n_0$ is the birefringence of a solvent, $\Delta\tau$ is the shear stress, $g$ is the shear rate. Since the gap was small, the shear rate was considered constant.

Intrinsic viscosities of the polymer solutions have been measured with Ostwald's capillary viscometers with standard methods.

## Results and discussion

### Equilibrium and non-equilibrium electric birefringence

Fig. 2 illustrates the equilibrium electric birefringence data for solutions of cylindrical dendrimers.

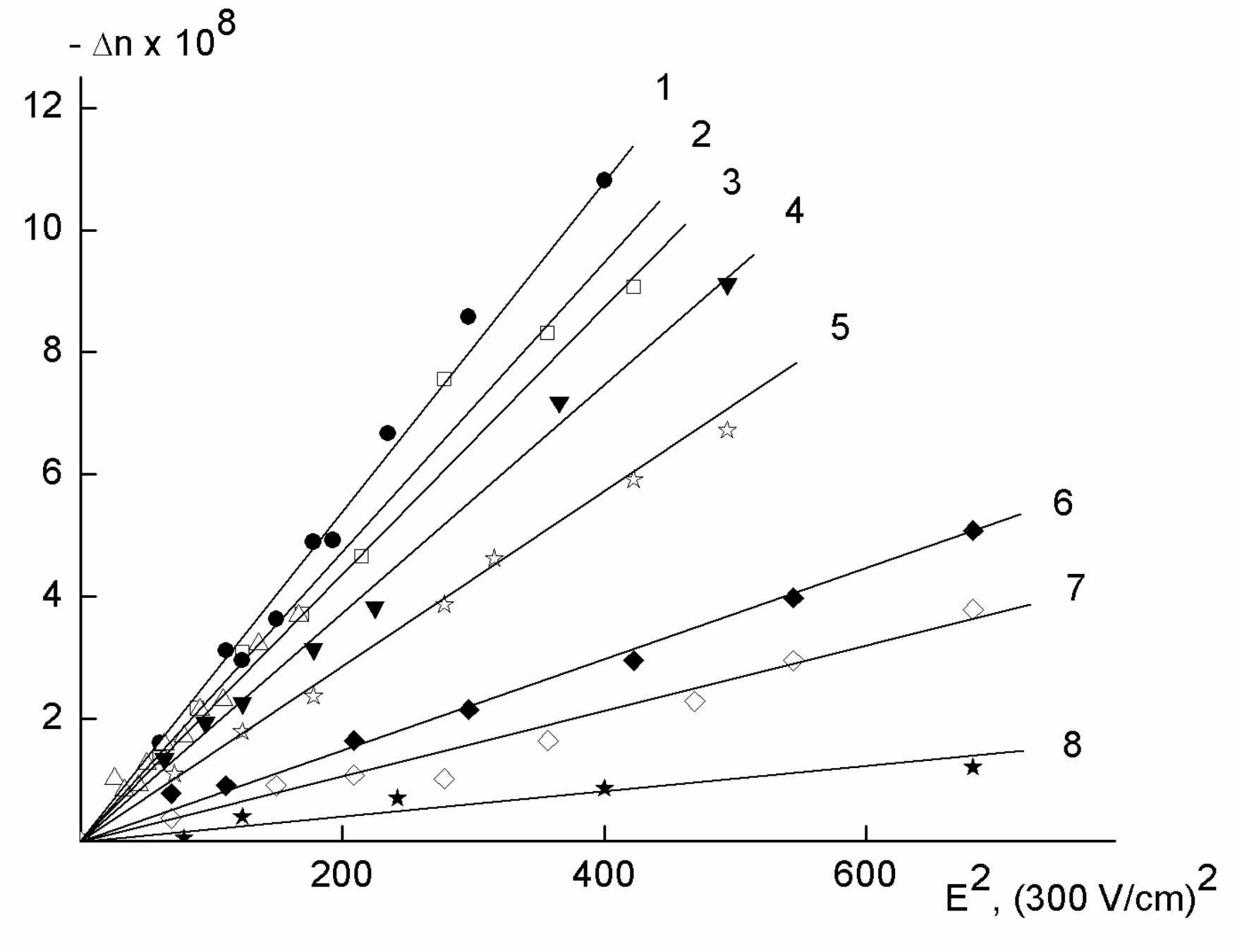

Fig. 2. Dependences of birefringence $\Delta n$ on the square of rectangular-pulsed electric field $E^2$ for the chloroform solution of P3-3 sample at the mass concentrations $c = 2.88 \cdot 10^{-2}$ $g/cm^3$ (1), $c = 1.67 \cdot 10^{-2}$ $g/cm^3$ (2), $c = 1.10 \cdot 10^{-2}$ $g/cm^3$ (3), $c = 0.75 \cdot 10^{-2}$ $g/cm^3$ (4), $c = 0.52 \cdot 10^{-2}$ $g/cm^3$ (5), $c = 0.17 \cdot 10^{-2}$ $g/cm^3$ (6), $c = 0.09 \cdot 10^{-2}$ $g/cm^3$ (7), and data of pure chloroform (8).

In the studied strength range, these obtained plots are straight lines obeying the Kerr law. Analogous relationships were observed for other samples at various concentrations. Fig. 3 shows the concentration dependences of the equilibrium Kerr constant for several samples.

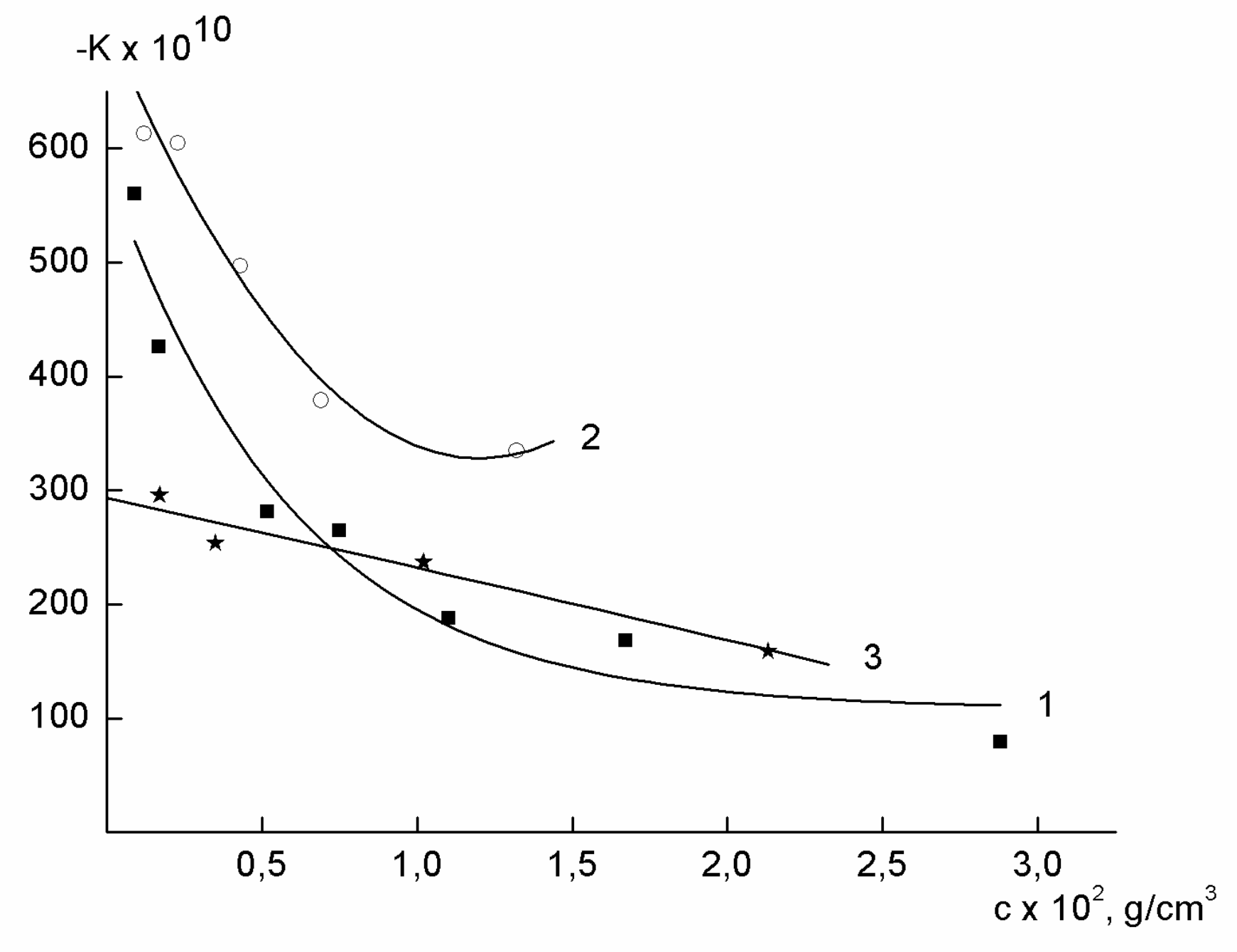

Fig. 3. Dependence of Kerr constants on the mass concentration of polymers P3-3 (1), P3-4 (2), P3-5 (3).

The absolute values of Kerr constants drastically increase as a solution is diluted. Such concentration dependences have been earlier observed for various polymers, such as aromatic polyamides solved in dimethylamide and dimethylsulphoxide. The growth is most likely caused by charge effects and the change of dielectric properties of solutions at dilution. Chloroform, in which the samples have been studied, is a much less polar and less conducting solvent than dimethylamide and dimethylsulphoxide. However, it is necessary to emphasize that the number of highly polar amide groups per unit chain length in studied polymers is seriously higher than in the aromatic polyamides. This fact leads to such strong concentration dependences of Kerr constants. While electrooptical properties of solutions demonstrate the discussed unusual concentration behavior, the concentration dependences of all other parameters (hydrodynamic and optical) have quite usual form.

The results obtained with methods of non-equilibrium electric birefringence are illustrated with Fig. 4 that shows the dependences of birefringence $\Delta n$ on the squared electric field $E^2$, for the sinusoidal-pulsed fields at different frequencies. Analogous dependencies have been obtained for every studied sample at varied concentrations.

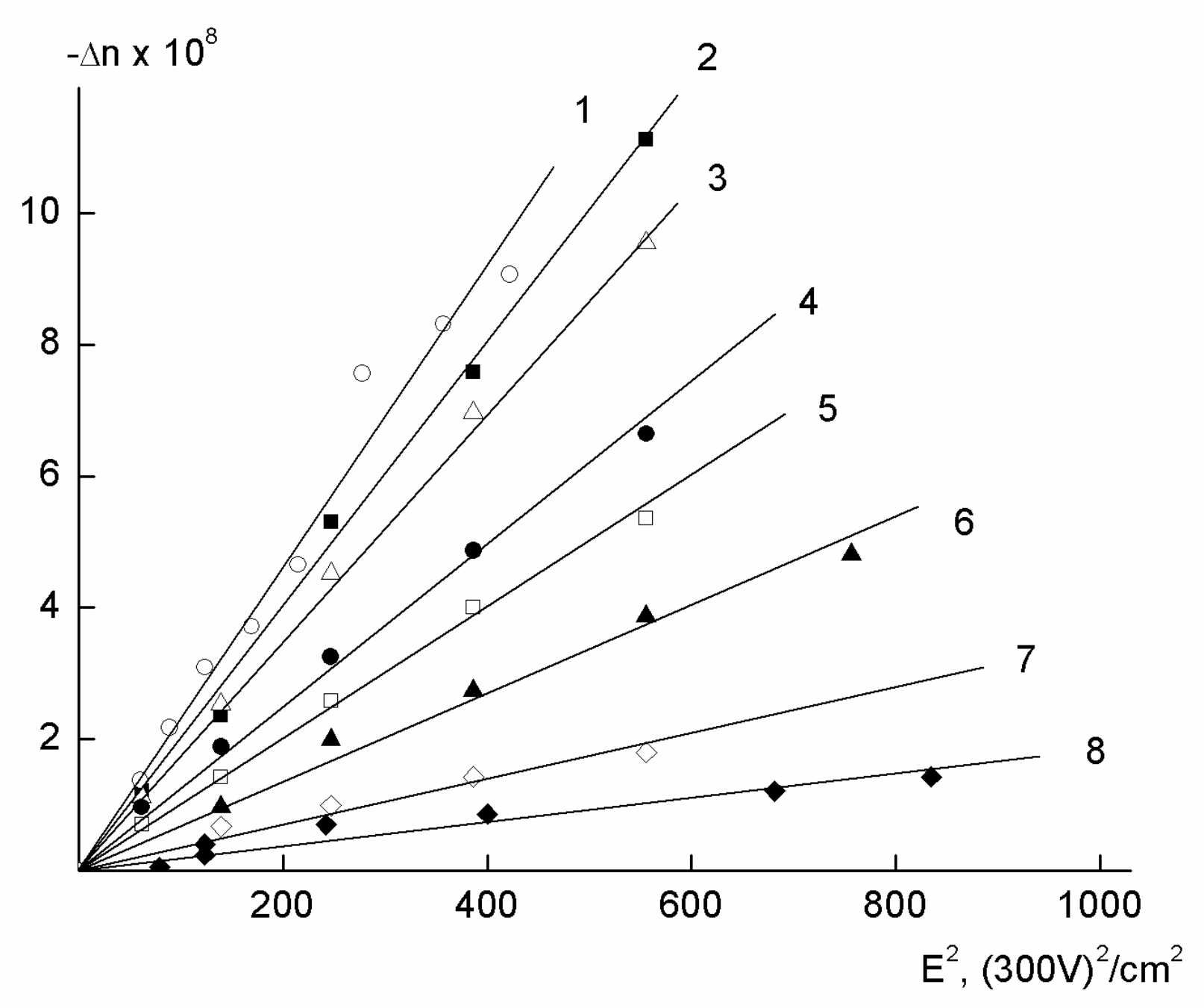

Fig. 4. Dependences of birefringence *Δn* on the square of sinusoidal-pulsed electric field $E^2$ for the chloroform solution of P3-3 sample at the mass concentration $c = 2.88 \cdot 10^{-2}\ g/cm^3$. The plot includes data for the rectangular-pulsed (*0 Hz*) electric field (1), sinusoidal fields with frequencies of *13 kHz* (2), *21 kHz* (3), *60 kHz* (4), *100 kHz* (5), *200 kHz* (6), *600 kHz* (7), and data of pure chloroform (8).

The specific Kerr constants were calculated using the said data. Fig. 5 displays the dispersion curves for a sample at varied concentrations. Fig. 6 shows the dispersion curves for different samples at a minimum researched concentration.

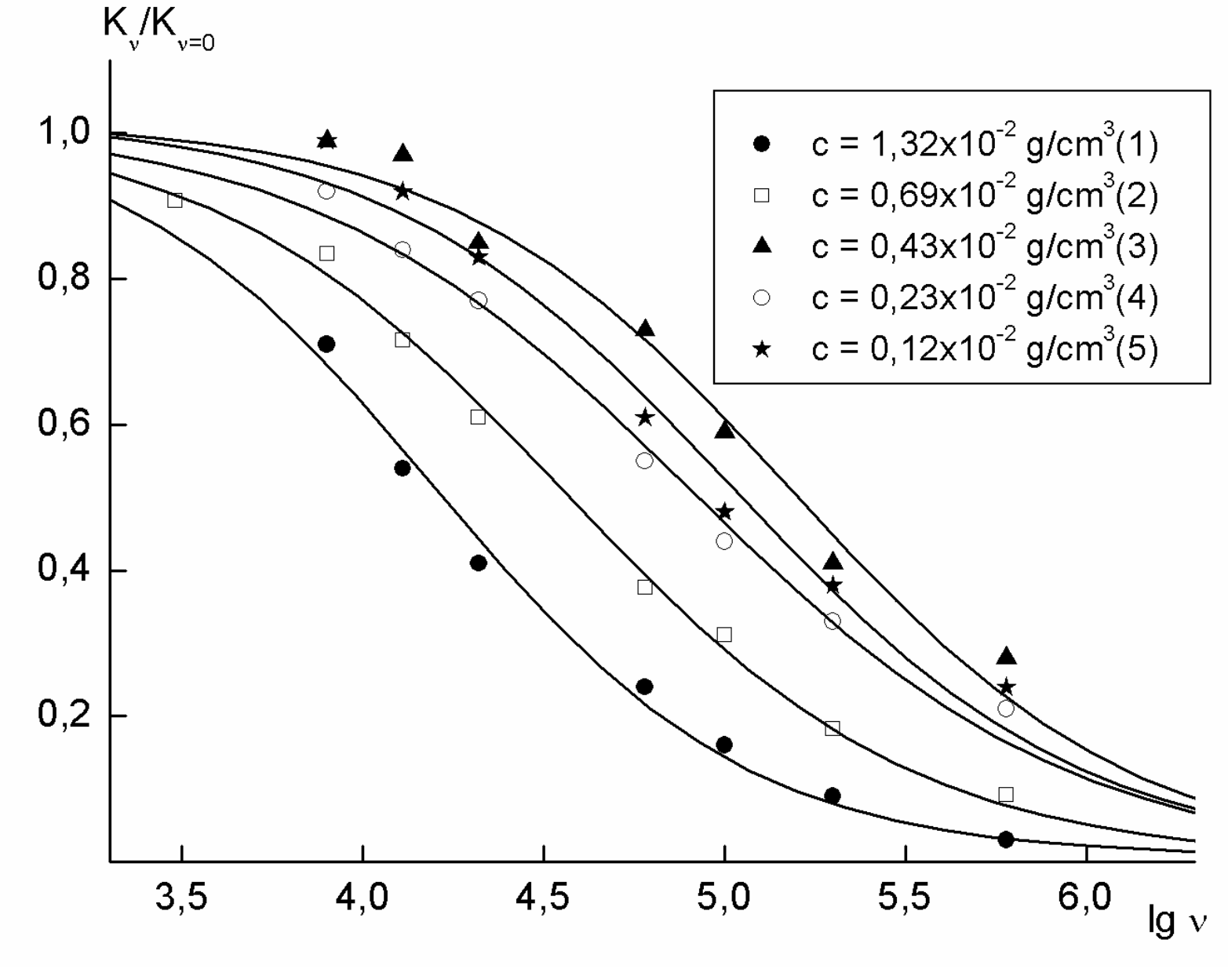

Fig. 5. Dispersion curves of specific Kerr constant $K_\nu/K_{\nu\to 0}$ for the chloroform solutions of P3-5 sample at mass concentrations $c = 1.32 \cdot 10^{-2}\ g/cm^3$ (1), $c = 0.69 \cdot 10^{-2}\ g/cm^3$ (2), $c = 0.43 \cdot 10^{-2}\ g/cm^3$ (3), $c = 0.23 \cdot 10^{-2}\ g/cm^3$ (4), $c = 0.12 \cdot 10^{-2}\ g/cm^3$ (5).

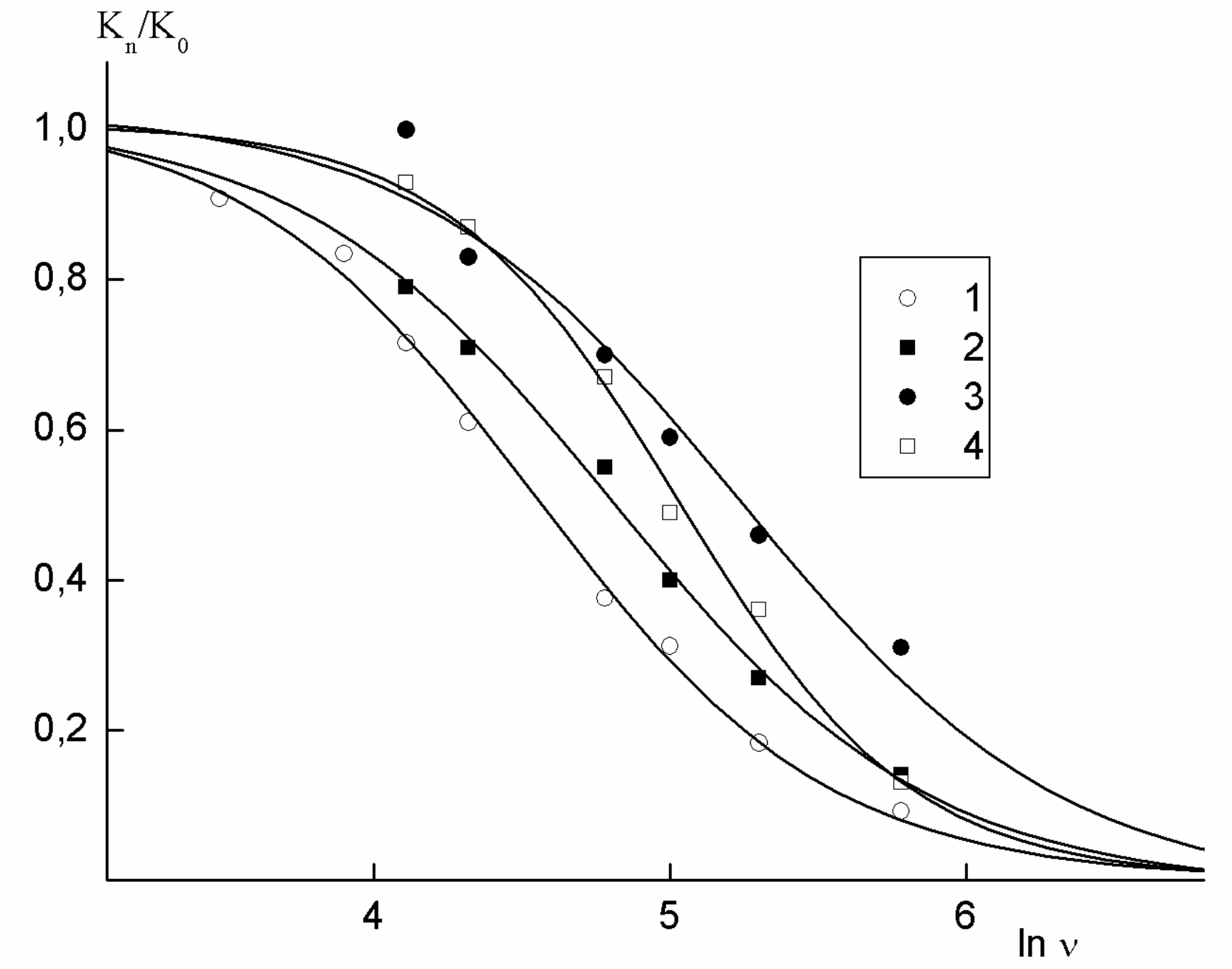

Fig. 6. Dispersion curves of specific Kerr constant $K_\nu/K_{\nu\to 0}$ for the chloroform solutions of polymers P3-5 (1), P3-3 (2), P3-1 (3), P3-2 (4), at the least studied concentrations.

Dispersion dependences of the samples decay virtually to zero which indicates that the macromolecules undergo the reorientation in external fields with the dipole mechanism, and the electric birefringence in solutions is explained by the constant dipole moments of the molecules. The mean dispersion relaxation times $\tau=1/(2D_r)$, where $D_r$ is the rotational diffusion coefficient relative to the short axis of a macromolecule, were found using the experimental curves. The expression to find the mean relaxation times is given by:

$$\tau = \frac{1}{\omega_n},$$

where $\omega_n=2\pi\nu_m$ is a frequency that corresponds to the half-decay of the electric birefringence dispersion effect:

$$K_{\nu_m} = \frac{K_{\nu=0} - K_{\nu\to\infty}}{2}.$$

Relaxation time $\tau=1/(2D_r)$, molar mass $M$ and intrinsic viscosity $[\eta]$ have a relation expressed by:

$$M[\eta]\eta_0 D_r=FRT,$$

where $D_r$ is a rotational diffusion coefficient with respect to short axis of a molecule, $F$ is a model parameter that characterizes the dimensions and the conformation of a molecule. For rigid molecules the values of $F$ range from $0.13$ for a rod to $0.42$ for a spherical particle [16].

The values of $F$ for molecules in chloroform solutions were in the interval *0.01...0.11*, which corresponds to a rigid rod. Therefore in chloroform the dendrimers have proven to follow the *large-scale reorientation*.

Analogous measurements of electric birefringence at different frequencies of electric field were performed for solutions of samples in dichloroacetic acid. It was found that the magnitude of electric birefringence decays with growing frequency $\nu$ of sinusoidal-pulsed field. However, the birefringence does not approach the half its equilibrium values at the frequencies below *1 Mhz*, which is the maximum possible value for the experimental equipment (Fig. 7).

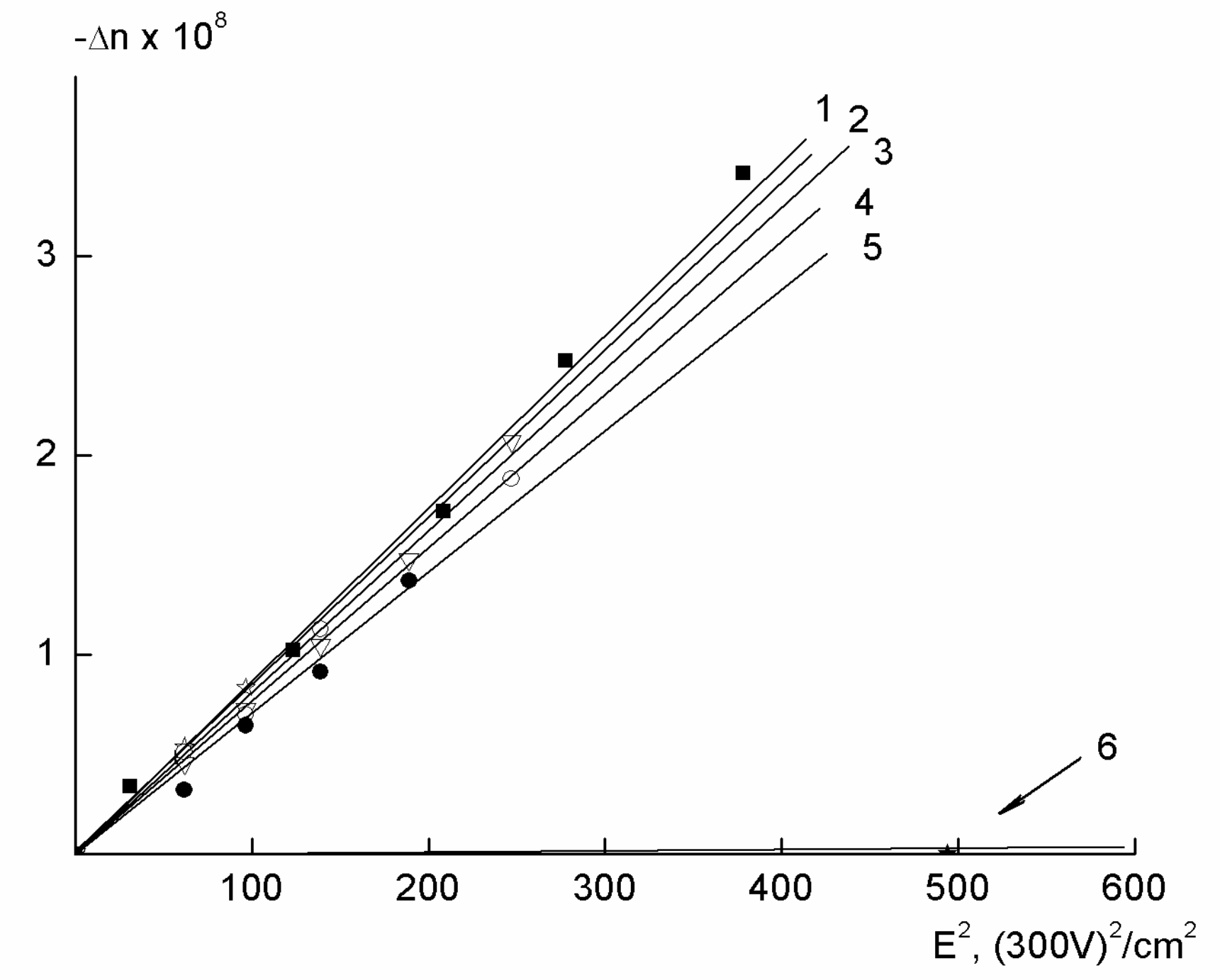

Fig. 7. Dependences of birefringence $\Delta n$ on the square of sinusoidal-pulsed electric field $E^2$ for the dichloroacetic acid solution of P3-1 sample at the mass concentration $c = 0.84{\cdot}10^{-2}$ $g/cm^3$. The plot includes data for the rectangular-pulsed (*0 Hz*) electric field (1), sinusoidal fields with frequencies of *20 kHz* (2), *100 kHz* (3), *200 kHz* (4), *550 kHz* (5), and data of pure dichloroacetic acid (6).

The upper limit of the relaxation times and, respectfully, the lower limit of the coefficient $F$ may be estimated if the first points of the dispersion dependence are extrapolated by the Debye curve. The values of $F$ for studied samples in DCA are in the interval *3...6* which seriously exceeds the corresponding values for chloroform solutions and the theoretically possible value *0.42* for kinetically rigid molecules. This fact allows to make a conclusion that the reorientation of particles in DCA in electric fields follows the large-scale mechanism because the relaxation times corresponding to such large coefficients $F$ are impossibly small for rigid particles.

This result may be explained by degradation of hydrogen bonds and consequent growth of kinetic flexibility of polymer chains in DCA. The values of $F$ directly indicate the low-scale mechanism of reorientation of molecules in electric field. An analogous behavior was earlier observed for cylindrical dendrimers of lower generations with dendrons based on L-aspargic acid [12], [13].

Fig. 8 illustrates the measured flow birefringence in dependence of the applied shear stress $\Delta\tau=g(\eta-\eta_0)$ for one of the dedrimers at two different concentrations. All studied samples have shown analogous linear dependences of birefringence on shear stress $\Delta\tau$ and no influence of concentration on the slope.

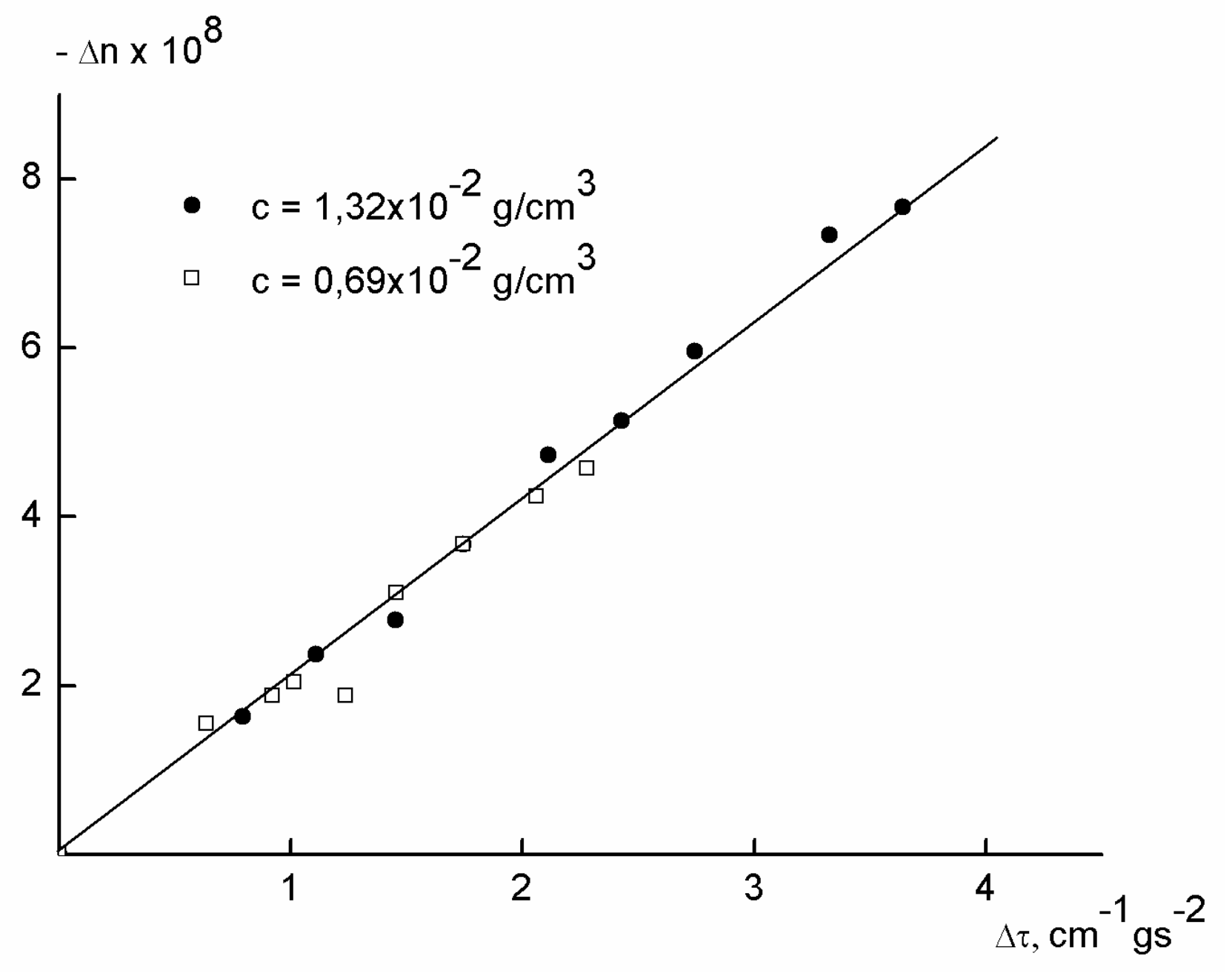

Fig. 8. Dependence of the birefringence $\Delta n$ on the shear stress $\Delta\tau$ for the dichloroacetic acid solutions of P3-5 samples at mass concentrations $c = 1.32\cdot10^{-2}\ g/cm^3$ (1) and $c = 0.69\cdot10^{-2}\ g/cm^3$ (2).

The calculated Kerr constants $(-100...-300\times10^{-10}\ g^{-1}cm^5(300\ V)^{-2})$ and flow birefringence constants $(-50...-200\times10^{-2}\ s^2cm/g)$ exceed the values obtained for dendrimers of lower generations.

## Theoretical models for the shape of particles

The conformational analysis of the studied macromolecules was performed on the basis of the values for translational coefficients of diffusion found experimentally in cooperation with the Institute of Macromolecular Compounds, Russian Academy of Sciences (St Petersburg) by the method of isothermal diffusion [17].

According to the theory developed by Fujita et al. [18, 19], the translational friction factor $f$ for a spherocylinder at statistical orientation is equal to:

$$f = \frac{3\pi\eta_0 L}{\ln L/d + Q(d/L)},$$

where $Q(d/L)$ is the power series for the axial ratio $d/L$ given by:

$$Q\left(\frac{d}{L}\right) = 0{,}3863 + 0{,}6863 \cdot \left(\frac{d}{L}\right) - 0{,}06250 \cdot \left(\frac{d}{L}\right)^2 - 0{,}01042 \cdot \left(\frac{d}{L}\right)^3 -$$
$$- 0{,}000651 \cdot \left(\frac{d}{L}\right)^4 + 0{,}0005859 \cdot \left(\frac{d}{L}\right)^5 + \ldots$$

In this expression, $L$ is the particle's length, $d$ is its cross-section size (diameter), $\eta_0$ is the solvent viscosity. The spherocylindrical shape is used to model the studied macromolecules due to the determined fact that lengths and diameters of particles are comparable [17].

The experimental translational coefficients of diffusion were used to calculate the values of translational coefficients of friction $f$ of a single particle. The lengths of the polymer chains were found from the polymerization degrees and the known monomer unit projection on the main polymer chain.

Approximation of $f$ in dependence of the length of a polymer chain $L$ has shown that the Fujita formulas cannot describe the experimental data at any reasonable values of the diameter of molecules (Fig. 9).

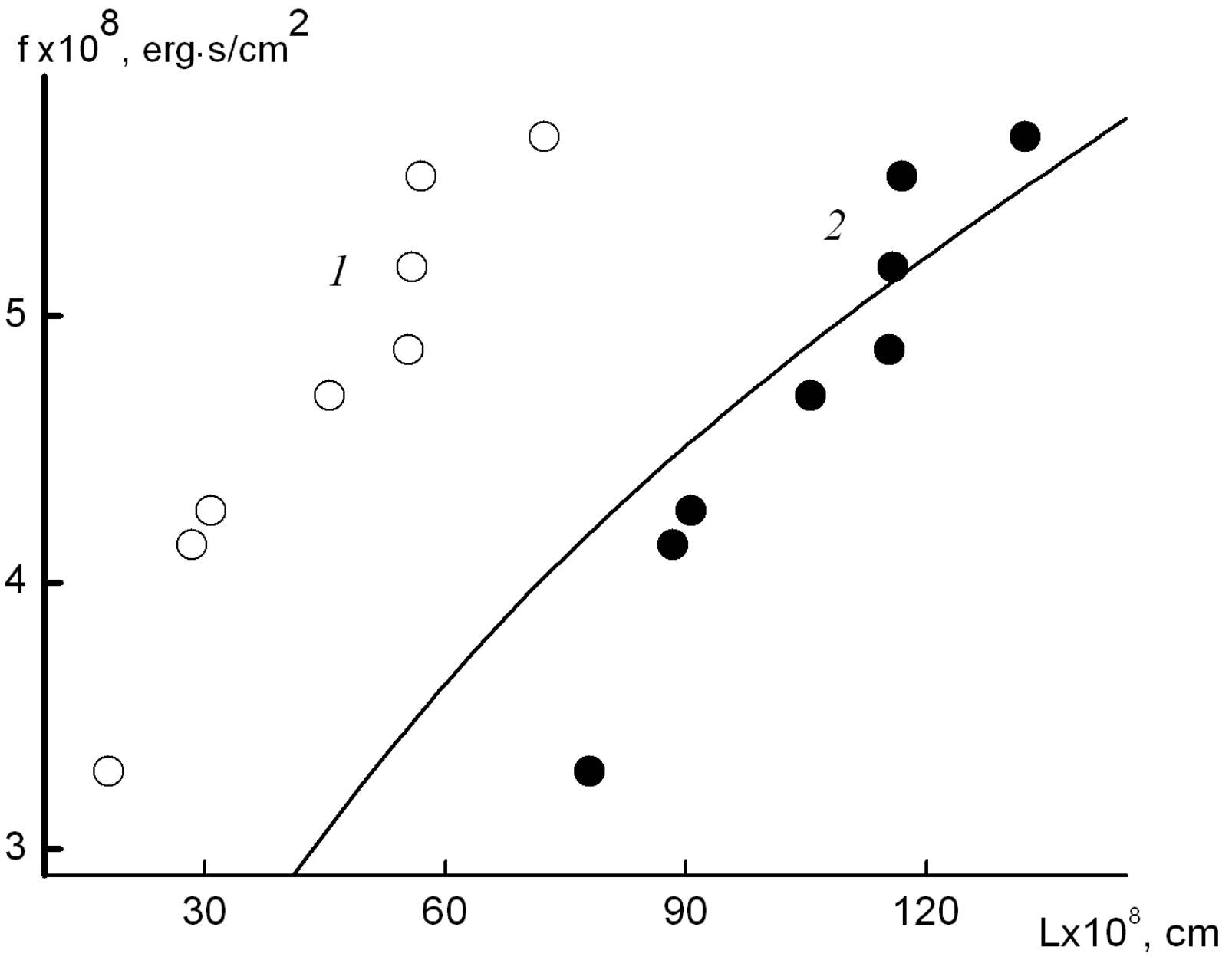

Fig. 9. Dependence of the translational friction factors on the length of molecules. (1) is the data for the calculated lengths of main polymer chains, (2) is the same data increased with *60 Å* and reflecting the real lengths of molecules with terminal dendrons oriented along the polymer chain. The theoretical curve shows the best fit for rigid spherocyliners.

However, if all experimental data is displaced *60 Å* to the right, a satisfying correlation between theory and experiment is achieved (Fig. 9). This fact is evidence that the linear dimensions of macromolecules are exterior of the corresponding lengths of polymer chains.

This observation may be interpreted in the following way. Spatial repulsive intermolecular interactions force terminal dendrons (attached to "first" and "last" segments of main polymer chain) to be oriented along the chain, not normally to it.

Diameters of molecules were estimated as *80 Å*, which makes the observed elongation by *60 Å* quite reasonable.

This orientation of terminal dendrons leads to the increase of the hydrodynamic radii for the polymers of least molar masses, and consequently, to maximum deviation of corresponding experimental points from the experimental curve.

## Conclusions

The Kerr and the flow birefringence constants for all studied samples exceed the corresponding values for the samples of first and second generations. This fact may be related to increased equilibrium rigidity and increased number of anisotropic groups per unit length of polymer chain.

Dispersion dependences show that the macromolecules undergo reorientation in external electric fields because of their constant dipole moments. The mechanism of reorientation is strongly dependent on the physical and chemical properties of the solvent.

In chloroform solutions, the studied macromolecules align to the microwave-frequency electric fields according to large-scale mechanism. In other words, carbochain polymers acquire high kinetic rigidity when side dendritic substituents are attached to the main chain. However, for dichloroacetic acid solutions, the macromolecules undergo reorientation according to low-scale mechanism.

The difference in the reorientation mechanism is explained by degradation of intermolecular hydrogen bonds caused by molecules of dichloroacetic acid.

Analysis of the translational coefficients of friction led to a grounded assumption that the terminal dendritic substituents are oriented mainly along the primary polymer chain. This assumption is in accordance to the results obtained with methods of electric and flow birefringence.